\documentclass[12pt]{article}
\pdfoutput=1
    \newcommand{\be}[1]{\begin{equation}\label{#1}}
    \newcommand{\ba}[1]{\begin{eqnarray}\label{#1}}
    \newcommand{\ep}[1]{\epsilon_{#1}}
    \newcommand{\ga}[1]{\gamma_{#1}}
    \newcommand{\de}[1]{\delta_{#1}}

    \newcommand{\pa}[1]{\left(#1\right)}

    \newcommand{\M}{{\rm M_{\rm P}}}

    \def\ee{\end{equation}}
    \def\ea{\end{eqnarray}}

\usepackage{graphicx,latexsym}
\usepackage{graphicx,epsf}
\usepackage{color}
\usepackage{epsfig}
\usepackage{amsmath}
\usepackage[T1]{fontenc}
\usepackage[utf8]{inputenc}
\usepackage{tikz}
\usepackage{pgfplots}
\usepackage{authblk}
\begin{document}
\title{Asymptotically Safe Non-Minimal Inflation}

\author[1]{Alessandro Tronconi\thanks{Alessandro.Tronconi@bo.infn.it}}
\affil[1]{Dipartimento di Fisica e Astronomia and INFN, Via Irnerio 46,40126 Bologna,
Italy}
\date{}
\maketitle

\begin{abstract}
We study the constraints imposed by the requirement of Asymptotic Safety on a class of inflationary models with an inflaton field non-minimally coupled to the Ricci scalar. The critical surface in the space of theories is determined by the improved renormalization group flow which takes into account quantum corrections beyond the one loop approximation. The combination of constraints deriving from Planck observations and those from theory puts severe bounds on the values of the parameters of the model and predicts a quite large tensor to scalar ratio. We finally comment on the dependence of the results on the definition of the infrared energy scale which parametrises the running on the critical surface.
\end{abstract}
\section{Introduction}
Scalar-Tensor (ST) theories of gravitation have attracted much attention in recent years. Such theories naturally emerge from various attempts of quantising gravity and may thus play a major role in the description of the early Universe. In Scalar-Tensor theories the propagation of the scalar degree of freedom affects the value of the Newton's constant, possibly leading to observable deviations from Standard General Relativity (GR) predictions which could then be measurable through Solar System observations. Inflation \cite{inflation} in the context of Scalar-Tensor theories has been intensively studied. It has been shown that the spectra of cosmological perturbations generated during inflation \cite{pert} are frame independent i.e. they can be calculated either in the defining frame, dubbed Jordan frame (JF),  or in the conformally related Einstein frame (EF). While it is widely accepted that the equivalence holds at the classical level (at least in the absence of ordinary matter), at the quantum level it is not \cite{EJframe}.\\
Non-minimally coupled theories are particular Scalar-Tensor theories with the scalar field having a canonical kinetic term and a non-minimal coupling with the gravitational sector (in particular with the Ricci scalar). Such a coupling is always generated by the quantum corrections to the scalar field itself \cite{indgrav} and it is thus worth considering how it effects the dynamics of standard GR at planckian energies. Since the Higgs boson discovery \cite{higgs} the scenario of inflation consisting of a non-minimally coupled Higgs field, the first elementary scalar particle observed so far, has been suggested and investigated in detail \cite{higgsinf}. Non-minimal Higgs inflation successfully generates the observed spectra of primordial perturbations and is compatible with the Standard Model constraints at much lower energy. It takes place in the so-called induced gravity (IG) regime (where the Planck mass is negligible w.r.t. the non-minimal coupling term) and it is driven by the quartic self interaction of the Higgs field. In such a regime, the Higgs potential gives a slow rolling (SR) evolution and a quasi de Sitter phase. The observational constraints coming from the amplitude of scalar perturbations combined with those on the self-coupling derived from the measured Higgs mass require a very large value for the non-minimal coupling. From the phenomenological point of view, Higgs inflation (and its variants such as the Starobinsky model which can be obtained by simply redefining the dynamical degrees of freedom) is among the inflationary models most compatible with latest Planck data \cite{planck}. However the stability of the Higgs potential with respect to quantum corrections has been highly debated and such a problem (called the ``eta problem'') needs to be overcome in general for a satisfactory inflationary model building. \\
General Relativity is the classical theory for the description of gravitational interactions. Despite its success in describing the long distance interactions GR must be considered as an effective theory when quantum effects associated with the gravitational field are taken into account. As a quantum theory GR is perturbatively non-renormalizable, namely the UV divergences generated at all orders in perturbation theory need, in principle, an infinite number of counter-terms and of measurements to fix the free parameters of the theory. This is not a problem if GR is treated as an effective description of gravity at energy well below the Planck scale. At these energies the counter-terms are suppressed by inverse powers of the Planck mass and their effect can be neglected. However, at planckian energies, the quantum corrections becomes important and the framework loses predictivity. Because of this, at least in the description of the very early universe, a consistent theory of quantum gravity is needed.\\
In the last years the Asymptotic Safety (AS) programme has been applied to the gravitational sector (also in the presence of matter fields) in order to overcome the problems mentioned above and provide a complete and consistent description of gravity at planckian energies \cite{AS}. The possible existence of a non trivial (interacting) UV fixed point for the gravity-matter system solves the problem related to the lack of predictivity mentioned above. If a fixed point exists it defines the so-called critical surface, a finite dimensional hyper-surface embedded in the space of theories whose points evolve into the fixed point in the UV. Given the hypothesis that the critical surface must contain the quantum theory which describes our universe, then this latter theory is complete in the UV by construction, in the sense that its high energy behaviour is controlled by the fixed point and a finite number of observables (equal to the dimension of the critical surface) are sufficient to specify the UV description of nature.\\ 
On a less rigorous ground compared to its original formulation based on the Wilson RG ideas, the AS programme has recently been intensively used for cosmological model building by connecting the free parameters of the resulting models to the  inflationary and the late time cosmology observables \cite{alfio}. \\
The UV completion of a set of Scalar-Tensor theories has been analysed in the AS framework \cite{ASgippo} and the non-trivial fixed points of a theory in $d$ dimensions in the presence of $N$ scalar fields with $O(N)$ symmetry has been studied. Close to the fixed point the RG flow generates the shape of the potential of the scalar field and that of its coupling to gravity. For $N=1$ the result is quite simple: the scalar field is non-minimally coupled to gravity and its potential contains a cosmological constant, a mass term and quartic self-interaction. These are the dynamical settings we shall investigate in this paper.\\

The article is organised as follows. In Section 2 we review the relevant equations governing the improved RG flow, their fixed points and the linearised theory around one of these points. 
In Section 3 we discuss the formalism for both the homogeneous sector (for two relevant potentials  derived from the RG flow) and the primordial perturbations. Let us note that all the details of the derivation sketched in sections 2 and 3 can be found in the Ref. \cite{ASgippo}. We then derive the expressions for inflationary observables. In Section 4 we
illustrate the general constraints on the space of parameters coming from the CMB observations. In Section 5 we discuss the same models with respect to the theoretical constraints imposed by the improved RG flow and finally in Section 6 we draw the conclusions.   
\section{Improved RG formalism}
We start from the general action for a ST theory with one scalar field $\phi$
\be{genform}
S=\int d^4x \pa{U(\phi)R-\frac{1}{2}g^{\mu\nu}\partial_\mu \phi\partial_\nu \phi-V(\phi)}
\ee
where $U$ and $V$ are generic functions of $\phi$. These functions are subjected to a RG flow parametrised by the cutoff $k$. The non-linear differential equations which govern the flow are derived in Ref. \cite{ASgippo} and can be conveniently written in terms of the dimensionless quantities $v\equiv V/k^4$, $u\equiv U/k^2$ and $\varphi\equiv \phi/k$ as
\ba{flowvfull}
\dot v \!\!\!\!&=&\!\!\!\!-4 v+ \varphi\,  v' +c_4 \frac{9}{2}+
c_4 \frac{ 5 \left( \dot u- \varphi  u'\right)}{6 u}\nonumber\\
&{}&\!\!\!\!+c_4\frac{6 u+ \left(1\!+\!v''\right) \left( \dot u\!-\!3 u\!- \varphi  u'\right)\!+\!
3 u' \left(2 \dot u'\!+\!3 u'\!-\!2 \varphi  u''\right)}{6 \left[3 \left(u'\right)^2+ u \left(1+v''\right)\right]}
\ea
\ba{flowufull}
\dot u \!\!\!\!&=&\!\!\!\!-2 u+ \varphi  u'+c_4\frac{157}{36}+c_4\frac{5 \left( \dot u- \varphi  u'\right)}{9\, u} -c_4 \left( u u''+ u'{}^2\right) \times \nonumber\\
&{}&\!\!\!\!\times \frac{2 u^2+u'{}^2 \left( \varphi  u' - \dot u\right)+2 u u' \left(- \varphi  u''+3 u'+
   \dot u'\right)}{u \left[3 \left(u'\right)^2+ u \left(v''+1\right)\right]^2} 
   \nonumber\\
   &{}&\!\!\!\!+c_4 \frac{  \varphi u' \left[6 u''+ \left(v''+1\right)\right]-6 u'\dot u'+  4\, u v''-\dot u
   \left(v''+1\right)}{12 \left[3 \left(u'\right)^2+ u \left(v''+1\right)\right]}
\ea
where the prime denotes the derivative w.r.t. $\varphi$, the dot is the derivative w.r.t. the RG time ($\tau\equiv \log k$) and $c_4=1/32\pi^2$. Let us note that the above equations are obtained by studying the exact flow equation for the effective average action, first of all truncating the action in terms of local operators according to two functions. This choice corresponds to the so called ``derivative expansion of the effective action''. Within such an expansion the wave function renormalisation and operators that are powers of curvatures and derivatives thereof can be neglected as they contribute to NLO effects. In particular the inclusion of these latter NLO contribution would introduce several complications at practical (for example inflationary dynamics would be driven by two scalar fields in the Einstein frame) and at more fundamental level. This approach still corresponds to the choice of a truncation with an infinite number of couplings. The derivation is based on the background field method where an exponential parametrisation of the metric has been adopted (see \cite{ASgippo} for details) and  a ``type I'', spectrally adjusted, cutoff has been introduced.\\
The equations (\ref{flowvfull}), (\ref{flowufull}) admit 2 non trivial fixed points. The fixed point solution FP1 does not depend on $\varphi$ and is given by
\be{fp1}
u_{\rm FP}=\frac{157}{2304\pi^2}\sim 7\cdot 10^{-3},\;v_{\rm FP}=\frac{5}{128\pi^2}\sim4\cdot 10^{-3}.
\ee
Let us note that the values (\ref{fp1}) are small and of the same order of magnitude. Physically the bare theory at FP1 corresponds to GR plus a positive cosmological constant.\\
A second fixed point solution with $v$ constant and $u$ proportional to $\varphi^2$ can be found and has the following form
\be{fp2}
u_{\rm FP}=-\frac{41}{420}\varphi^2\sim - 10^{-1}\, \varphi^2,\;v_{\rm FP}=\frac{3}{128\pi^2}\sim2\cdot 10^{-3}.
\ee
This latter fixed point solution has the ``wrong'' sign in front of the Ricci scalar and therefore violates the stability condition of the ST theory \cite{staro}. We then exclude it from the analysis which follows. Let us note that the adoption of the ``improved'' RG flow equations instead of the one loop, unimproved, ones reduces the number of fixed point solutions which can be found starting from the simplest assumptions on their functional form. If one generalises the above construction to the case of $N>1$ scalar fields with $O(N)$ symmetry other ``physically acceptable'' fixed points appear.
\subsection{Stability Analysis}
The linearisation of the equations (\ref{flowvfull}), (\ref{flowufull}) around FP1 gives important informations about the stability of the fixed point, relevant and irrelevant directions of the theory and the shape of a small patch of the critical surface around FP1. If one expands around FP1 
\be{expFP1}
u=u_{\rm FP}+ \delta u,\; v=v_{\rm FP}+ \delta v
\ee
the equations (\ref{flowvfull}), (\ref{flowufull}) take the following form
\ba{lineq}
\dot {\delta v}\!\!\!\!\!&=&\!\!\!\!\!\frac{5}{32\pi^2}-4v_{\rm FP}-4\delta v+\frac{\dot{\delta u}}{32\pi^2 u_{\rm FP}}-\frac{\varphi\,\delta u'}{32\pi^2u_{\rm FP}}+\varphi\,\delta v'-\frac{\delta v''}{32\pi^2},\label{lineq1}\\
\dot {\delta u}\!\!\!\!\!&=&\!\!\!\!\! \frac{157}{1152\pi^2}-2u_{\rm FP}-2\delta u+\frac{17\,\dot {\delta u}}{1152\pi^2u_{\rm FP}}+\varphi\,\delta u'-\frac{17\varphi\,\delta u'}{1152\pi^2u_{\rm FP}}\nonumber\\
&&\!\!\!\!\!-\frac{\delta u''}{32\pi^2}+\frac{\delta v''}{96\pi^2}.\label{lineq2}
\ea
The above linearised equations are a good approximation near the fixed point. While the concept of ``distance'' in the space of theories would require a long discussion, on observing the form of Eqs. (\ref{lineq1}) and (\ref{lineq2}) we argue that the linear approximation works pretty well if $|\delta v|$ and $|\delta u|/u_{\rm FP}$ (and their derivatives) are small enough. Let us note that the condition $|\delta v|\ll 1$ is less restrictive than $|\delta u|/u_{\rm FP}\ll 1$ since $v_{\rm FP}$ and $u_{\rm FP}$ are much smaller than one. If they are both satisfied we conclude that $(u,v)$ is closed to the fixed point.\\
On solving Eqs. (\ref{lineq1}) and (\ref{lineq2}) one finds the following critical exponents with the corresponding eigenvectors
\ba{crexp}
&\!\!\!\!\!\!\!\!\!\!\!\!\!\!\!\!\!\theta_1=4, & \!\!\!\!\!\!\!\   (\delta v,\delta u)_1 = (1,0)\label{eig1} \\
&\!\!\!\!\!\!\!\!\!\!\!\!\!\theta_2=\frac{314}{123}, & \!\!\!\!\!\!\!\ 
(\delta v,\delta u)_2 = \left(-\frac{72}{89}, 1\right) \label{eig2}\\
&\!\!\!\!\!\!\!\!\!\!\!\!\!\!\!\!\!\theta_3=2, &  \!\!\!\!\!\!\!\ 
(\delta v,\delta u)_3 = 
\left(\varphi^2-\frac{29}{544\pi^2},
\frac{157}{3264\pi^2}\right)\label{eig3}\\
&\!\!\!\!\!\!\!\!\!\!\!\!\!\theta_4=\frac{68}{123},& \!\!\!\!\!\!\!\ 
(\delta v,\delta u)_4
=\bigg(-\frac{72}{89}\varphi^2+\frac{56913}{3094352\pi^2},
\varphi^2-\frac{17741}
{350304\pi^2}\bigg)\label{eig4}
\\
&\!\!\!\!\!\!\!\!\!\!\!\!\!\!\!\!\!\theta_5=0,& \!\!\!\!\!\!\!\  (\delta v,\delta u)_5 
=\bigg(\varphi^4-\frac{87}{272\pi ^2}\varphi^2+\frac{87}{17408\pi^4},
\frac{471 }{1632\pi^2}\varphi^2-\frac{645}{52224\pi ^4}\bigg)\label{eig5}.
\ea
We conclude that there are four relevant directions and one marginal direction. Close to FP1 the RG flow spans the critical surface implicitly defined by
\be{crsur}
(v,u)(t)=(v_{\rm FP},u_{\rm FP})+\sum_{i=1}^5 c_i(\delta v,\delta u)_i{\rm e}^{-\theta_i t}\equiv (v_{\rm FP}+\delta v_t,u_{\rm FP}+\delta u_t)
\ee
where $c_i$ are arbitrary integration constants. 
\section{Inflationary dynamics}
The expression (\ref{crsur}), given the form of the eigeinvectors (\ref{eig1}-\ref{eig5}), generates the following functional dependence for $V$ and $U$:
\be{renpot}
V_{\rm RG}=\Lambda+\frac{m_2}{2}\phi^2+\frac{\lambda}{4}\phi^4
\ee 
and
\be{rencoup}
U_{\rm RG}=\frac{M_2+\xi\phi^2}{2}.
\ee
where the coefficients $\Lambda$, $m_2$, $\lambda$, $M_2$ and $\xi$ may be positive, negative or negligible. Since we are interested in viable inflationary models we shall restrict our analysis to two particular choices of these parameters.\\
Let us note that, for realistic inflationary models, the scalar field potential must be non negative and negligible at the minimum of the potential ($V(\phi_m)\simeq 0$). Correspondingly one can either have a potential $V_{1}$ with one minimum and
\be{c1}
\Lambda=0,\;m_2>0,\; \lambda >0
\ee
with $\phi_m=0$ or a potential $V_2$ with two minima for
\be{c2}
\Lambda=\frac{m_2^2}{4\lambda},\;m_2<0,\; \lambda >0
\ee
with $\phi_m=\pm\sqrt{-\frac{m_2}{\lambda}}$.
In the former case the potential has a symmetric ground state while in the latter the symmetry is spontaneously broken. We shall refer to $V_1$ as the symmetric case and to $V_2$ as the symmetry breaking case.\\ 
Furthermore we restrict our analysis to the case $\xi>0$ and $M_2>0$ (henceforth $M_2\equiv \M^2$). The overall sign of $U$ must be positive. Given the freedom of choosing the initial conditions for the inflaton field, negative values of $\xi$ and $M_2$  may lead to a negative ``effective'' gravitational constant and gravitons with a negative energy. The condition $U>0$ is also known as the stability condition for the ST theory.\\
If we restrict the analysis to the homogeneous dynamics and consider the usual flat FRW metric 
\be{metric}
g_{\mu\nu}={\rm diag}\pa{-1,a(t)^2,a(t)^2,a(t)^2}
\ee 
we obtain the following Friedmann equation
\be{fr0}
3H^2=\frac{1}{\M^2+\xi\phi^2}\pa{\frac{1}{2}\dot \phi^2+V-6\xi H \phi\dot \phi},
\ee
with $H\equiv{\dot a}/a$ and the dot means the derivative w.r.t. cosmic time. The Klein-Gordon equation can be cast in the following form
\be{kg1}
\ddot \phi+3H\dot \phi+\frac{\xi\pa{1+6\xi}\phi^2}{\M^2+\xi\pa{1+6\xi}\phi^2}\frac{\dot \phi^2}{\phi}+V_{{\rm eff},\phi}=0
\ee 
where the effective potential is implicitly defined by
\be{dVeff}
V_{{\rm eff},\phi}\equiv\frac{\pa{\M^2+\xi\phi^2}V_{,\phi}-4\xi\ \phi V}{\M^2+\xi\pa{1+6\xi}\phi^2}.
\ee
The dynamics of the scale factor and that of the scalar field during the inflationary period can be conveniently described by the slow roll (SR) parameters. One may then introduce the Hubble flow functions hierarchy $\ep{0}\equiv H_0/H$, $\ep{i+1}\equiv\frac{\dot {\ep{i}}}{H\ep{i}}$ and the hierarchy related to the evolution of the scalar field $\de{0}\equiv \phi/\phi_0$, $\de{i+1}\equiv\frac{\dot {\de{i}}}{H\de{i}}$. These two hierarchies are related by the equations of motion and, in particular, the second Friedmann equation is
\be{ep1}
\ep{1}=\frac{\frac{1+4\xi}{2\xi}\de{1}^2+\de{1}\pa{\de{2}-1}}{\rho+\pa{1+\de{1}}}
\ee
where 
\be{rhodef}
\rho\equiv \frac{\M^2}{\xi \phi^2}=\frac{\M^2}{\xi \phi_0^2\de{0}^2}.
\ee
The value of $\rho$, compared with unity, discriminates between the regime with a negligible coupling ($\rho\gg 1$) and that with a dominant coupling ($\rho\ll 1$). In the former case the dynamics is very close to GR while in the latter it is similar to IG. \\
It is convenient to introduce the dimensionless parameter 
\be{alphadef}
\alpha\equiv\frac{\lambda}{\xi}\frac{\M^2}{m_2}\ge 0
\ee
which will frequently appear in the calculations. The homogeneous dynamics and that of the inhomogeneous part of the matter-gravity system can be studied in terms of the three dimensionless quantities $\alpha$, $\xi$ and $\rho$. Let us finally note that, in the limit $m_2\rightarrow 0$, the potentials $V_1$ and $V_2$ must coincide. Such a limit can be expressed in terms of $\alpha$ and $\xi$ as the combined limits $\alpha\rightarrow -\infty$ and $-\alpha \xi\rightarrow +\infty$. \\
\subsection{Cosmological perturbations}
The inflationary observables are related to the inhomogeneities generated during inflation. The equations governing the dynamics of the cosmological perturbations for general Scalar-Tensor theories have been studied in a series of papers \cite{pert}. For the case (\ref{genform}) we are considering, 
the gauge invariant scalar perturbation can be described, in the uniform curvature gauge, in terms of the Sasaki-Mukhanov variable $v_s$ defined as
\be{defv}
v_s\equiv z_s\frac{H}{\dot \phi} \delta \phi
\ee
with
\be{defz}
z_s(t)\equiv\frac{a\dot \phi}{H}\sqrt{Z}\;\; {\rm and}\;\;Z\equiv \frac{1+\frac{3\dot U^2}{\dot \phi^2 U}}{\pa{1+\frac{\dot U}{2HU}}^2}.
\ee
Such a variable satisfies the following, second order, differential equation
\be{eqscal}
v_s''+\pa{k^2-\frac{z_s''}{z_s}}v_s=0
\ee
where henceforth the prime denotes a derivative w.r.t. the conformal time $\eta$ ($a\, d\eta=dt$) and 
\be{ddzoz}
\frac{z_s''}{z_s}=\frac{1}{\eta^2}\frac{\pa{1-\ga{1}+\ga{2}-\ga{3}+\ga{4}}\pa{2+\ga{2}-\ga{3}+\ga{4}}}{\pa{1+\ga{1}}^2}.
\ee
The hierarchy of $\ga{i}$'s in the above expression is defined as follows
\be{defgi}
\ga{1}\equiv\frac{\dot H}{H^2},\,\ga{2}\equiv\frac{\ddot \phi}{H\dot \phi},\,\ga{3}\equiv\frac{\dot U}{2HU},\;\ga{4}\equiv\frac{\dot E}{2H E}
\ee
and $E\equiv 2U\pa{1+\frac{3\dot U^2}{\dot\phi^2U}}$. In particular, in the non-minimal coupling case, one finds the following relations among the hierarchies $\ep{i}$, $\de{i}$ and $\gamma_i$
\be{gi}
\ga{1}=-\ep{1},\; \ga{2}=\de{1}+\de{2}-\ep{1},\;\ga{3}=\frac{\de{1}}{1+\rho},\;\ga{4}=\frac{\pa{1+6\xi}\de{1}}{\rho+1+6\xi}.
\ee
The scalar spectral index of the scalar perturbations can be obtained analytically to the first order in SR and is 
\be{nsm1}
n_s-1=-2\pa{\ep{1,*}+\de{1,*}+\de{2,*}-\frac{\de{1_*}}{1+\rho_*}+\frac{\pa{1+6\xi}\de{1_*}}{\rho_*+1+6\xi}}
\ee
where the subscript $*$ indicates that the quantity should be evaluated at the time the pivot scale $k_*$ exits the horizon during inflation (about 60 e-folds before inflation ends).\\
In the tensor sector each perturbation $h_i$, with $i=+,\times$ describing the two independent polarizations, may be conveniently described by the variable $v_T\equiv z_T h_i$ which satisfies the differential equation
\be{tenseq}
v''_T+\pa{k^2-\frac{z''_T}{z_T}}v_T=0
\ee
where $z_T\equiv a\sqrt{2U}$. To the second order in SR one finds the following expression for the spectral index for the tensor perturbations:
\be{nt}
n_t=-\frac{\de{1}^2}{\xi}\frac{\rho_*+1+6\xi}{\pa{\rho_*+1}^2}.
\ee
The tensor to scalar ratio measures the abundance of primordial gravitational waves generated during inflation 
and is given by
\be{r}
r\equiv\frac{\mathcal{P}_{h}}{\mathcal{P}_{\mathcal{R}}}=8\frac{\rho_*+1+6\xi}{\xi\pa{\rho_*+1}^2}\de{1,*}^2=-8n_t
\ee
where $\mathcal{R}\equiv v_s/z_s$ and
\be{spectra}
\mathcal{P}_{\mathcal{R}}\equiv\frac{k^3}{2\pi^2}\left|\mathcal R\right|^2,\;\mathcal{P}_{h}\equiv\frac{2k^3}{\pi^2}\pa{\left|h_+\right|^2+\left|h_\times\right|^2}.
\ee
are the power spectra of the scalar and tensor perturbations respectively.\\
Finally one finds the following general expression for the amplitude of the primordial scalar perturbations:
\be{PR}
\mathcal{P}_{\mathcal{R}}=\frac{H_*^2}{4\pi^2}\frac{1}{\phi_*^2\de{1,*}^2}\frac{\pa{\rho_*+1+\de{1_*}}^2}{\pa{\rho_*+1}\pa{\rho_*+1+6\xi}}\simeq \frac{H_*^2}{4\pi^2}\frac{1}{\phi_*^2\de{1,*}^2}\frac{\rho_*+1}{\rho_*+1+6\xi}
\ee
which has the observed value $\mathcal{P}_{\mathcal{R}}^{\rm (obs)}\simeq 2.4\cdot 10^{-9}$. 
\section{Comparison with observations}
In this section we impose the constraints derived from the inflationary observables on the parameters of the models discussed so far. The analytic prediction for the scalar spectral index and the tensor to scalar ratio are here presented for particular limits only (the complete analysis of non-minimal inflation with (\ref{renpot}), (\ref{rencoup}) will be discussed in detail in a separate paper \cite{inprep}). For each model the numerical results are also illustrated in the corresponding figures for particular choices of the parameters. A numerical treatment is necessary in order to extend the analysis to the regions of the parameter space which cannot be treated analytically. \\
Let us note that the non minimal coupling $\xi$ plays a crucial role in the homogeneous dynamics of the inflaton field. In particular for $\xi$ small enough SR can always occur leading to viable inflation while, when $\xi$ is large, the same does not happen. In general for the potential $V_1$ it is useful to distinguish among four possible regimes with inflation taking place in the IG or in the GR phase with either the quadratic or the quartic part of the potential dominating the dynamics of the scalar field. Four possibilities will be also considered for the potential $V_2$. In this case inflation may occur in the IG or in the GR phase and in the large field (LF) or in the small field (SF) regime. 
\subsection{Symmetric case}
\begin{figure}[t!]
\centering
\epsfig{file=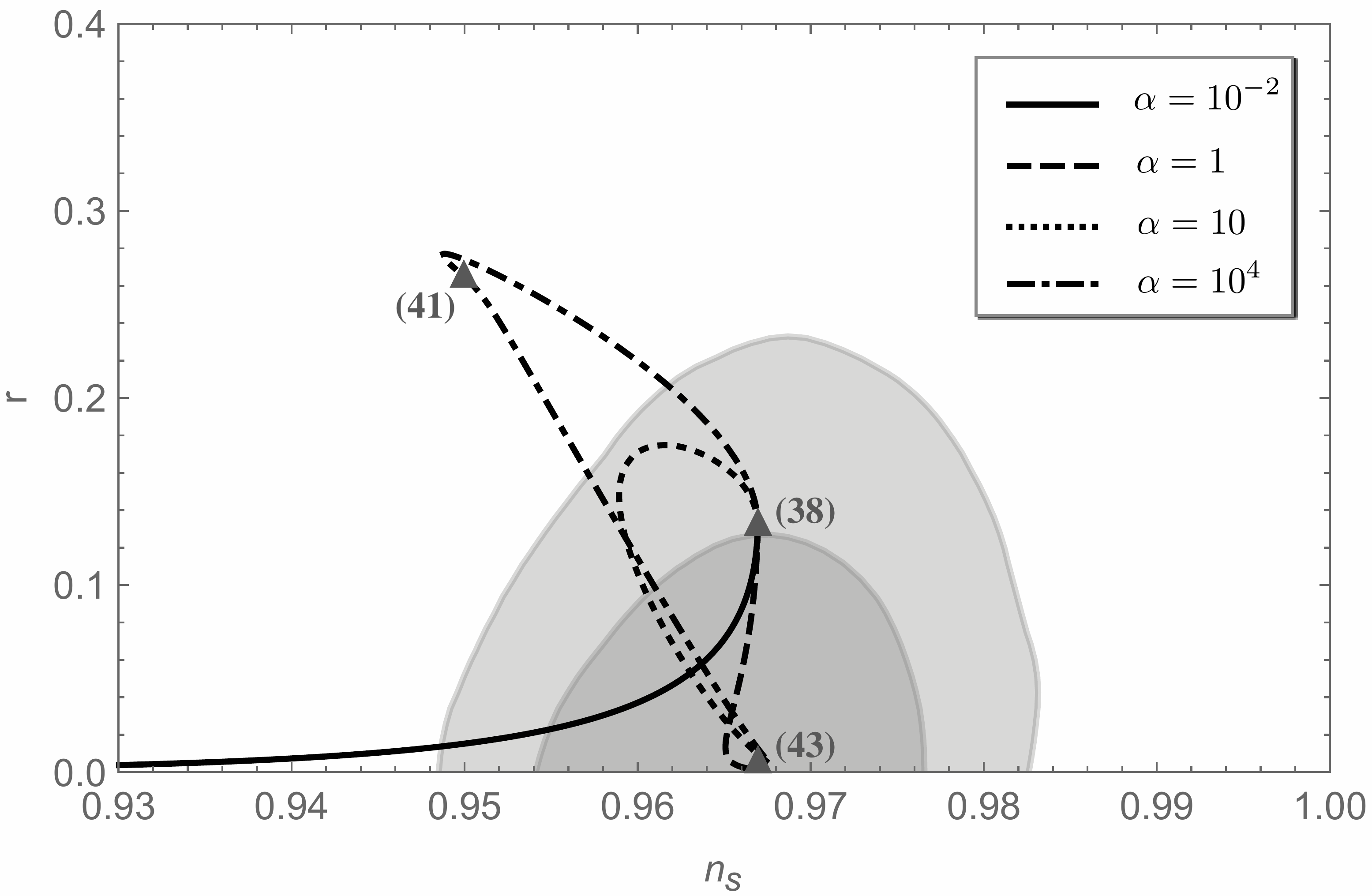, width=10. cm}
\caption{In this Figure we plotted the numerical predictions for $n_s$ and $r$ for the symmetric case. The four trajectories in each plot are obtained on varying $\xi$ and for $\alpha=10^{-2}$ (solid line), $\alpha=1$ (dashed line), $\alpha=10$ (dotted line) and $\alpha=10^4$ (dot-dashed line). The contours in the $(n_s,r)$ plane are the boundaries for the 68\% and 95\% confidence level regions obtained from {\it Planck 2015 TT+Low P} data. The grey triangles represent the points (\ref{PV1xs}), (\ref{PV1xsaxs}) and (\ref{PV1xlal}) for $N_*=60$.}
\label{fig1}
\end{figure}

In Figure (\ref{fig1}) we plotted the numerical values of $(n_s,r)$ with $N_*=60$, for different choices of $\alpha$ and letting $\xi$ vary from $10^{-15}$ to $10^{15}$. The values obtained are then compared with the observations from the latest Planck mission.  Let us note that all the trajectories exhibit a common behaviour. When $\xi\rightarrow 0$ they converge to the same point with coordinates
\be{PV1xs}
\pa{n_s,r}=\pa{1-\frac{2}{N_*},\frac{8}{N_*}},
\ee
while in the opposite, $\xi\gg 1$, limit $r\rightarrow 0$. \\
The coordinates (\ref{PV1xs}) can be found analytically for $\xi\ll 1$ and $\alpha \xi \ll 1$ and are exactly as expected since inflation occurs in the GR phase with the quadratic part of the potential dominating the dynamics. In GR, the large field models with monomial potentials of the form
\be{monpot}
V=\lambda\M^4\pa{\frac{\phi}{\M}}^n
\ee
give the following result
\be{nsrmon}
\pa{n_s,r}=\pa{1-2\frac{n+2}{4N_*+n},\frac{16n}{4N_*+n}}\stackrel{n=2}{\longrightarrow}\pa{1-\frac{2}{N_*},\frac{8}{N_*}}
\ee
which is reproduced by (\ref{PV1xs}).\\
For $\xi\ll 1$ but $\alpha \xi \gg 1$ inflation occurs in the GR regime with the quartic part of the potential dominating and one correctly finds
\be{PV1xsaxs}
\pa{n_s,r}=\pa{1-\frac{3}{N_*},\frac{16}{N_*}}.
\ee
This last point is outside the $95\%$ confidence level region in the $(n_s,r)$ plane of the Fig. (\ref{fig1}) and the trajectories obtained on varying $\xi$ and with $\alpha\gg 1$ pass close to it.\\
In the limit for $\xi\gg 1$ and $\alpha\ll 1$
\be{PV1xlas}
\pa{n_s,r}=\pa{-\frac{5}{3},0}
\ee
which is far from being compatible with observations (see Figure (\ref{fig1}) and the trajectory with $\alpha=10^{-2}$).\\
Finally when $\xi\gg 1$ and $\alpha \gg 1$
\be{PV1xlal}
\pa{n_s,r}\simeq\pa{1-\frac{2}{N_*},\frac{12}{N_*^2}}
\ee
i.e. one recovers the predictions of pure IG with a Landau-Ginzburg potential and a large coupling $\xi$, namely those of Higgs inflation.\\
Let us note that the two limits for $\xi$ large and small and a finite $\alpha$ are consistent with the following qualitative estimate: since $\alpha=\lambda \M^2/ \pa{\xi m_2}$ in order to keep $\alpha$ finite in the limit $\xi\rightarrow 0$ (when one recovers GR) one needs either $\lambda\ll 1$ or $m_2/\M^2 \gg 1$, namely the quadratic part of the potential dominates and one has chaotic inflation. On the other hand, for $\xi\gg 1$ the IG part dominates and in order to keep $\alpha$ finite one needs either $\lambda\gg 1$ or $m_2/\M^2\ll 1$ namely the quartic part of the potential is dominant.\\
The observed amplitude of the scalar perturbations constrains, through (\ref{PR}), the $m_2/\M^2$ ratio which take the following specific form
\be{massratioV1}
\frac{m_2}{\M^2}=48\pi^2 \mathcal{P}_{\mathcal{R}}^{\rm (obs)}\frac{\delta_{1,*}^2\pa{\rho_*+1+6\xi}}{\alpha\pa{1+2\frac{\rho_*}{\alpha}}}.
\ee
\begin{figure}[t!]
\centering
\epsfig{file=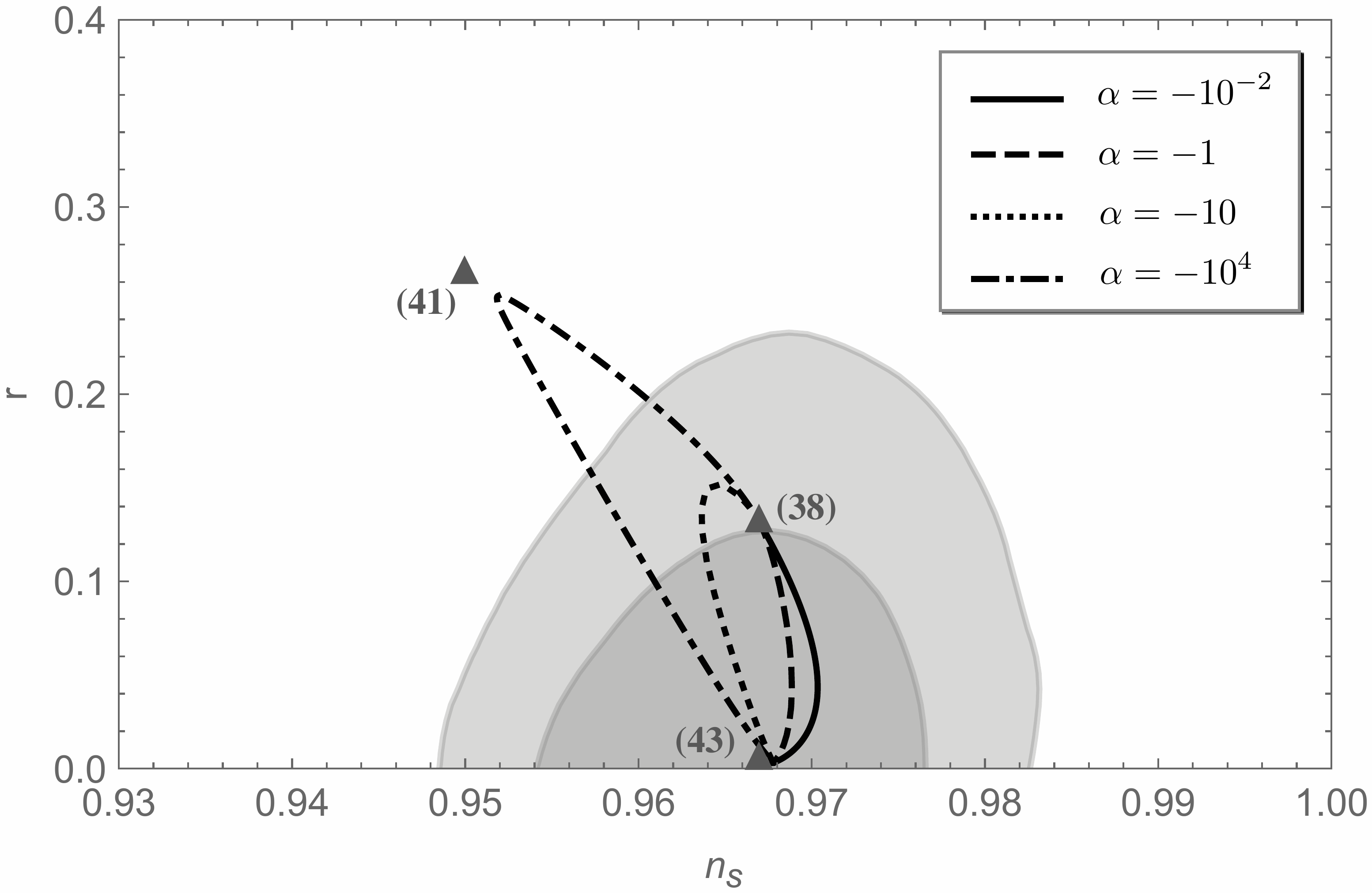, width=10. cm}
\caption{In this Figure we plotted the numerical predictions for $n_s$ and $r$ for the symmetry breaking (LF) case. The four trajectories in each plot are obtained on varying $\xi$ and for $\alpha=-10^{-2}$ (solid line), $\alpha=-1$ (dashed line), $\alpha=-10$ (dotted line) and $\alpha=-10^4$ (dot-dashed line). The contours in the $(n_s,r)$ plane are the boundaries for the 68\% and 95\% confidence level regions obtained from {\it Planck 2015 TT+Low P} data. The grey triangles represent the points (\ref{PV1xs}), (\ref{PV1xsaxs}) and (\ref{PV1xlal}) for $N_*=60$.}
\label{fig2}
\end{figure}
\subsection{Symmetry breaking case - Large Field}
In Figure (\ref{fig2}) the theoretical predictions for $(n_s,r)$ are compared with observations for $N_*=60$, different values of $\alpha$ and on letting $\xi$ vary for $10^{-15}$ to $10^{15}$. The trajectories exhibit a similar behaviour and, in contrast with the symmetric case, converge to the same two points in the limits $\xi\rightarrow 0$ and $\xi\rightarrow +\infty$.\\
Analytically for $\xi\rightarrow 0$ one finds the same limit as obtained for the symmetric case (\ref{PV1xs}). In the opposite limit, when $\xi\rightarrow \infty$, we obtain the result (\ref{PV1xlal}).\\ 
Finally, when $\xi\ll 1$ and $-\alpha\xi\gg 1$, the trajectories pass close to (\ref{PV1xsaxs})\\
One can finally constrain the ratio $-m_2/\M^2$ through (\ref{PR}) which now has the following form
\be{massratioV2LF}
\frac{m_2}{\M^2}=48\pi^2\mathcal{P}_{\mathcal{R}}^{\rm (obs)}\frac{\delta_{1,*}^2\pa{\rho_*+1+6\xi}}{\alpha\pa{1+\frac{\rho_*}{\alpha}}^2}.
\ee
\subsection{Symmetry breaking case - Small Field}
\begin{figure}[t!]
\centering
\epsfig{file=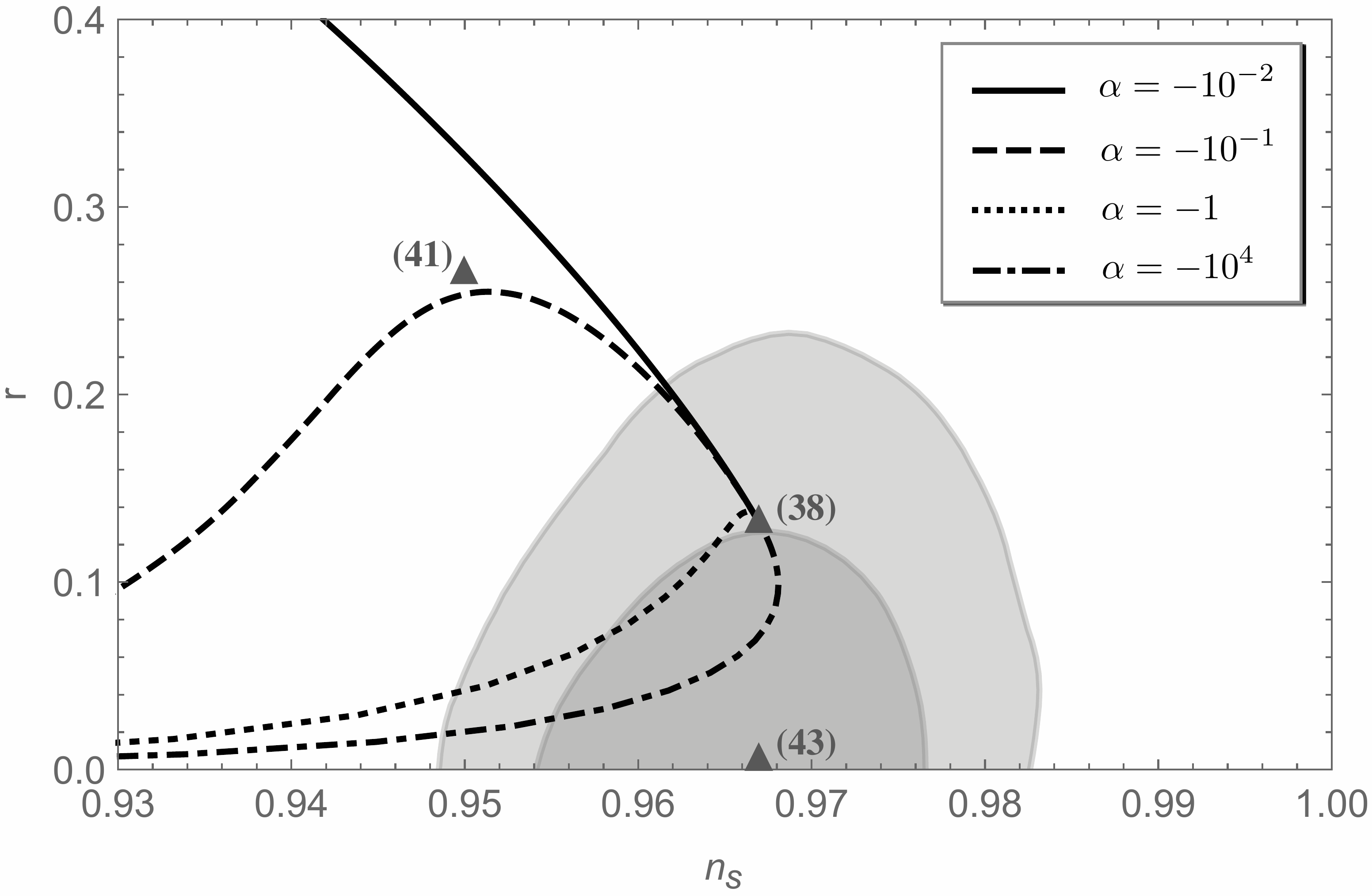, width=10. cm}
\caption{In this Figure we plotted the numerical predictions for $n_s$ and $r$ for the symmetry breaking case with the field increasing toward the minimum and different values of alpha. The four trajectories in each plot are obtained on varying $\xi$ and for $\alpha=-10^{-2}$ (solid line), $\alpha=-1$ (dashed line), $\alpha=-10$ (dotted line) and $\alpha=-10^4$ (dot-dashed line). The contours in the $(n_s,r)$ plot are the boundaries for the 68\% and 95\% confidence level regions obtained from {\it Planck 2015 TT+Low P} data. The grey triangles represent the points (\ref{PV1xs}), (\ref{PV1xsaxs}) and (\ref{PV1xlal}) for $N_*=60$.}
\label{fig3}
\end{figure}

In this last case we only discuss the $\xi\ll 1$ limit. Away from this limit the SR conditions do not hold. Figure (\ref{fig3}) compares the theoretical predictions for $(n_s,r)$ to observations for $N_*=60$, different values of $\alpha$ and on varying $\xi$ from $10^{-12}$ to $1$. The trajectories converge to (\ref{PV1xs}) when $\xi\rightarrow 0$ (or $-\alpha \xi\ll 1$). When $\xi\ll 1$ and $-\alpha \xi\gg 1$ then $\de{1}\simeq -4\xi \rho_*\gg1$ and SR is not possible.
One can finally constraint the ratio $-m_2/\M^2$ through the observed amplitude of the scalar perturbations and the expression (\ref{PR}).
\begin{figure}[t!]
\centering
\epsfig{file=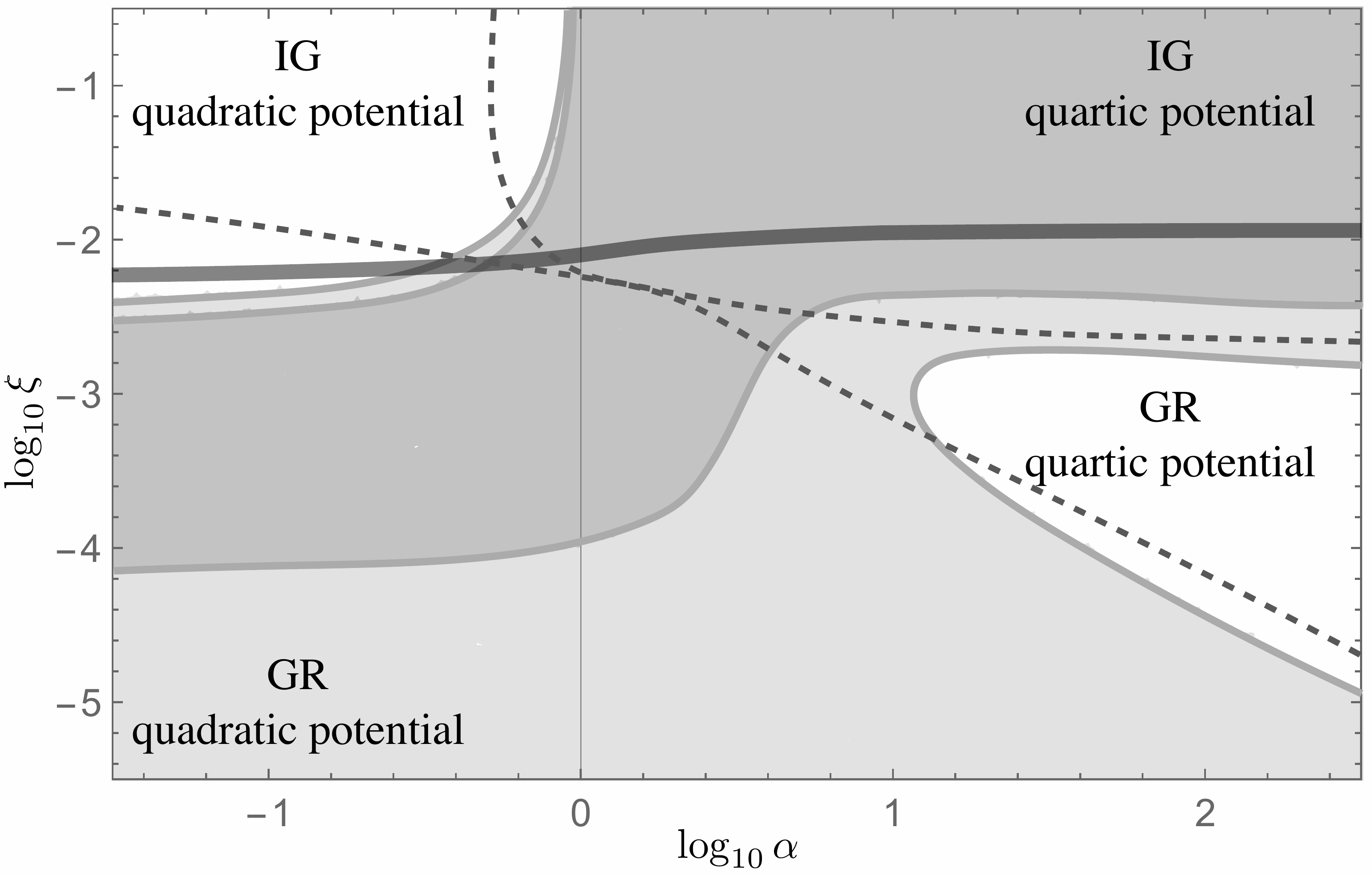, width=11. cm}
\caption{In this Figure, 3 different regions are shown. The darker grey region represents the set of $(\alpha,\xi)$ values leading to the theoretical predictions which lie inside the darker grey region in the $(n_s,r)$ plane of Figure (\ref{fig1}) (namely the $68\%$ confidence level region). Similarly the lighter grey region corresponds to the lighter grey part of the $(n_s,r)$ plane (namely the $95\%$ confidence level region) and finally the white area is that which leads to predictions outside the $95\%$ confidence level region. The dashed lines divide the parameters space in 4 regions corresponding to different inflationary regimes. The black band represents the subset of parameters corresponding to an evolution close to the fixed point (\ref{fp1}) for $k\sim \phi_*$.}
\label{fig1b}
\end{figure}
\section{RG Flow}
We have discussed, in the Introduction, at least qualitatively the motivations for restricting our analysis to a particular subset of scalar-tensor theories. The request that the viable inflationary models of the form (\ref{genform}) lie on the critical surface associated with the non trivial fixed point (\ref{fp1}) is a relevant constraint for model building. In order to verify if such a request is satisfied for some specific expressions for $U$ and $V$ at some scale $k$ one needs to study its RG flow in the $k\rightarrow \infty$ limit and verify that (\ref{fp1}) is the limit of such a flow. While the full set of equations which govern the RG flow are known (at least within certain approximations) and are given by (\ref{flowvfull}), (\ref{flowufull}), far from the fixed point (away from the linear regime) the expressions for $U$ and $V$ are a priori unknown. Then it is not an easy task to guess which models lie on the critical surface far from the fixed point in order to check their compatibility with inflationary observables. On the other hand for inflationary models near the fixed point the shape of $U$ and $V$ is determined by the linearised flow (\ref{lineq1},\ref{lineq2}) and one can find the values of their parameters compatible with observations, given the expressions listed in the previous sections. Let us note that one can generate models far from the fixed point on evolving with Eqs. (\ref{flowvfull}), (\ref{flowufull}) the theories which lie close to it down to some infrared scale and then check if the resulting theories are compatible with inflationary observations. This last approach seems less appropriate as it needs analytical inflationary predictions for a very general class of inflaton potentials and couplings to gravity.\\
In this section our aim is to choose, among the inflationary models which satisfy observational constraints, those lying on the critical surface close enough to the fixed point. These models are by definition asymptotically safe. If inflation takes place in the linear regime close to (\ref{fp1}) then the inflaton potential and its non-minimal coupling are those studied in this paper where, from the very beginning, the additional request of a negligible cosmological constant at the end of inflation was used.\\ 
Inflationary observables constrain $\xi$, $\alpha$ and the ratio $\mu\equiv m_2/\M$ while the value of the scalar field (or equivalently $\rho$) during inflation can be calculated from the equations of motion. Once the IR scale $k$, namely the energy scale of inflation, is set and for a given choice of $\xi$, $\alpha$ compatible with observation we are able to check if the selected model for inflation belongs to the critical surface (or at least lies very close to it). The rescaled inflaton potential, $v_*$ and non-minimal coupling, $u_*$, can be rewritten in terms of dimensionless quantities as
\be{rescaledV}
v_*=\frac{\Lambda}{k^4}+\frac{\M^4}{k^4}\frac{\mu}{2\xi\rho_*}+\frac{\M^4}{k^4}\frac{\alpha\mu}{4\xi\rho_*^2}
\ee
and
\be{rescaledF}
u_*=\frac{1}{2}\frac{\M^2}{k^2}\pa{1+\frac{1}{\rho_*}}
\ee
where $\Lambda/k^4$ is constrained by the potentials we are considering ($\Lambda=0$ for the symmetric case and $\Lambda=\frac{\mu \M^4}{4\alpha \xi}$ for the symmetry breaking case). The ratio $\M/k$ can be related to the identification of the infrared scale $k$. Such an identification has been much discussed in the literature \cite{AS,alfio}. Even if the RG formalism has solid grounds for the limits $k\rightarrow \infty$ and $k\rightarrow 0$, it is not clear how the dependence on $k$ in between should be physically interpreted. One is tempted to associate the IR scale with that of inflation. However in scalar-tensor theories (in contrast with GR where the energy scale of inflation is generally associated with $V_*^{1/4}$) this latter scale is not clearly defined.\\ 
In this section we shall consider two possibilities: first we shall associate $k$ with $\phi_*$, then we shall consider the identification $k\sim H_*$ which is more common in a cosmological context. In the first case one has
\be{Mpid1}
\frac{\phi_*^2}{k^2}=\frac{\M^2}{\xi \rho_*k^2}\simeq 1\Rightarrow \frac{\M^2}{k^2}\simeq \xi \rho_* 
\ee
while in the second case
\be{Mpid2}
\frac{H_*^2}{k^2}=\frac{v_*}{3\frac{\M^2}{k^2}\pa{1+\frac{1}{\rho_*}}}\simeq 1\Rightarrow \frac{\M^2}{k^2}\simeq \frac{v_*}{3\pa{1+\frac{1}{\rho_*}}}
\ee
Different identifications are also possible. As a rule of thumb one may set $k$ to be close to one of the relevant mass scales in the model. However if one chooses $k\sim V_*^{1/4}$ or $k\sim U_*^{1/2}$ then the request that $(v_*,u_*)$ be close to the fixed point $(v_{FP},u_{FP})$ cannot be satisfied since $v_{FP}$ and $u_{FP}$ are much less than 1.
In contrast with other (not so uncommon) approaches in the literature \cite{AS,alfio} we shall not consider any dynamical implication associated with the above identification. \\
We already observed that, from the Eqs. (\ref{lineq1},\ref{lineq2}), one can argue that the linearisation procedure holds if 
\be{linearcond}
\left|v_{FP}-v_*\right|\ll 1\;{\rm and}\;\left|\frac{u_{FP}-u_*}{u_{FP}}\right|\ll 1
\ee 
and if such conditions are verified we can conclude that inflation takes place in the linear regime. Let us note that technically we shall look for models which verify the condition 
\be{linearcondt}
\left|v_{FP}-v_*\right|\le 10^{-2}\;{\rm and}\;\left|\frac{u_{FP}-u_*}{u_{FP}}\right|\le 10^{-2}
\ee 
and we neglect the systematic uncertainty associated with the exact identification of the IR scale.\\
An alternative method of verifying the above statement consists of comparing the r.h.s. of the Eqs. (\ref{flowvfull}), (\ref{flowufull}) and that of the linearised Eqs. (\ref{lineq1},\ref{lineq2}) evaluated on the linear solution (\ref{crsur}). The result of this comparison confirms that (\ref{linearcond}) is indeed a correct assumption.\\
Let us finally conclude by observing that all the constraints imposed so far (both from inflationary observables and from the theory) effect only dimensionless quantities. In order for some energy scale to appear in our predictions one needs to set the scale of some dimensional parameter in the theory. For example the requirement that, at the end of inflation, when the scalar field sits in the minimum of the potential $\phi_m$, the effective Planck mass is $\simeq 10^{19}\,{\rm GeV}$ would allow one to reconstruct all the energy scales in terms of it. 
\subsection{Symmetric case}
In this case the following $\M/k$ independent relation holds between $v$ and $u$:
\be{relfvV1}
v_*=\frac{2\mu}{\xi\rho_*}\frac{\pa{1+\frac{\alpha}{2\rho_*}}}{\pa{1+\frac{1}{\rho_*}}^2}u_*^2.
\ee
Given the form of (\ref{rescaledF}) we observe that $u_*\sim u_{FP}$ only for $\xi$ small. In such a case (on assuming $N_*=60$)  the linearity condition for $u$ constrains $\xi\sim 10^{-3}$ when $\alpha \xi\ll 1$ and $\xi\sim 10^{-2}$ when $\alpha \xi\gg 1$. In both cases $v_*$ satisfies the condition (\ref{linearcond}) since $\mu\ll 1$. We conclude that on identifying $k\sim\phi_*$ and imposing the linearity condition (\ref{linearcond}) we find a subset of asymptotically safe inflationary models compatible with observations. These models predict a tensor to scalar ratio $r$ of the order of the SR parameters. In contrast IG models with $\xi$ large predict a much smaller $r$ of the second order in SR.\\
On setting $k\sim H_*$, from the Friedmann equation one has $v_*\sim 6 u_*$ which in principle allows for both $v_*$ and $u_*$ to satisfy the linearity conditions. However for the symmetric case the relation (\ref{relfvV1}) combined with $v_*\sim 6 u_*$  leads to 
\be{fsV1}
u_*\sim \frac{1}{8\pi^2\mathcal{P}_{\mathcal R}^{(obs)}\de{1}^2}\frac{\xi\pa{\rho_*+1}^2}{\rho_*+1+6\xi}\gg 1
\ee
and we cannot conclude that $u_*$ is $\mathcal{O}\pa{u_{FP}}$.\\
In Figure (\ref{fig1b}) we plotted a relevant part of the parameter space and its comparison with observations.
In the figure, obtained by randomly spanning the parameter space and comparing the numerical outcome for $(n_s,r)$ to observations,  three regions are shown - a darker grey region, a lighter grey region and a white region - representing the choices of $(\alpha,\xi)$ which lead to a scalar spectral index and a tensor to scalar ratio respectively inside the 68\% or inside the 95\% or outside the 95\% confidence level region of the $(n_s,r)$ plane represented in Figure (\ref{fig1}). The parameter space is divided by a dashed line into 4 regions representing, on varying $\alpha$ and $\xi$ the possible regimes for which inflation takes place. The black band represents the subset of theories which satisfy the linearity conditions (\ref{linearcond}) on setting $k\sim \phi_*$.\\
\begin{figure}[t!]
\centering
\epsfig{file=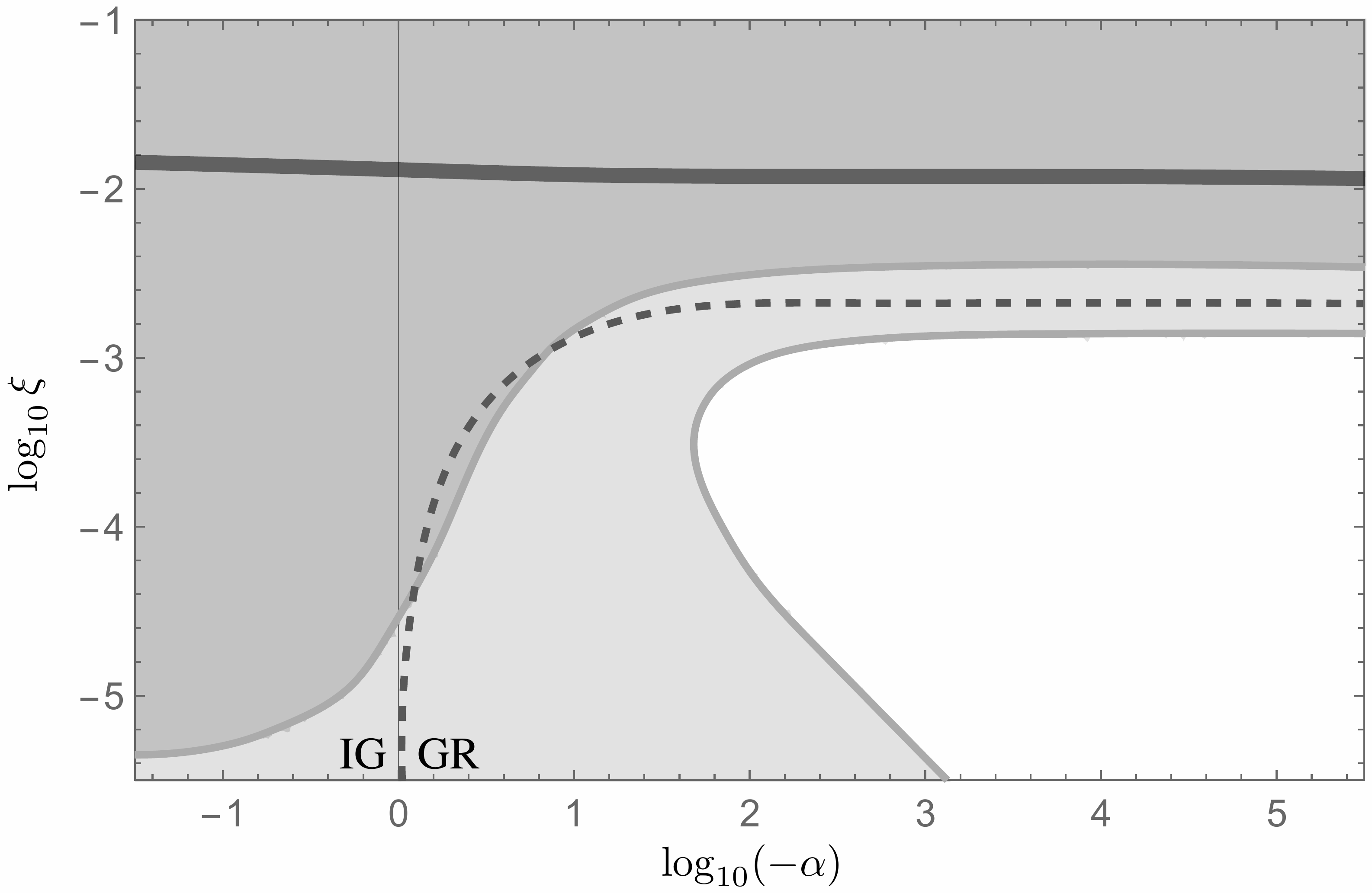, width=11. cm}
\caption{In this Figure, 3 different regions are shown. The darker grey region represents the set of $(\alpha,\xi)$ values leading to the theoretical predictions which lie inside the darker grey region in the $(n_s,r)$ plane of Figure (\ref{fig2}) (namely the $68\%$ confidence level region). Similarly the lighter grey region corresponds to the lighter grey part of the $(n_s,r)$ plane (namely the $95\%$ confidence level region) and finally the white area is that which leads to predictions outside the $95\%$ confidence level region. The dashed lines divide the parameters space in 2 regions corresponding to different inflationary regimes. The black band represents the subset of parameters corresponding to an evolution close to the fixed point (\ref{fp1}) for $k\sim \phi_*$.}
\label{fig2b}
\end{figure}
\subsection{Symmetry breaking case}
\begin{figure}[t!]
\centering
\epsfig{file=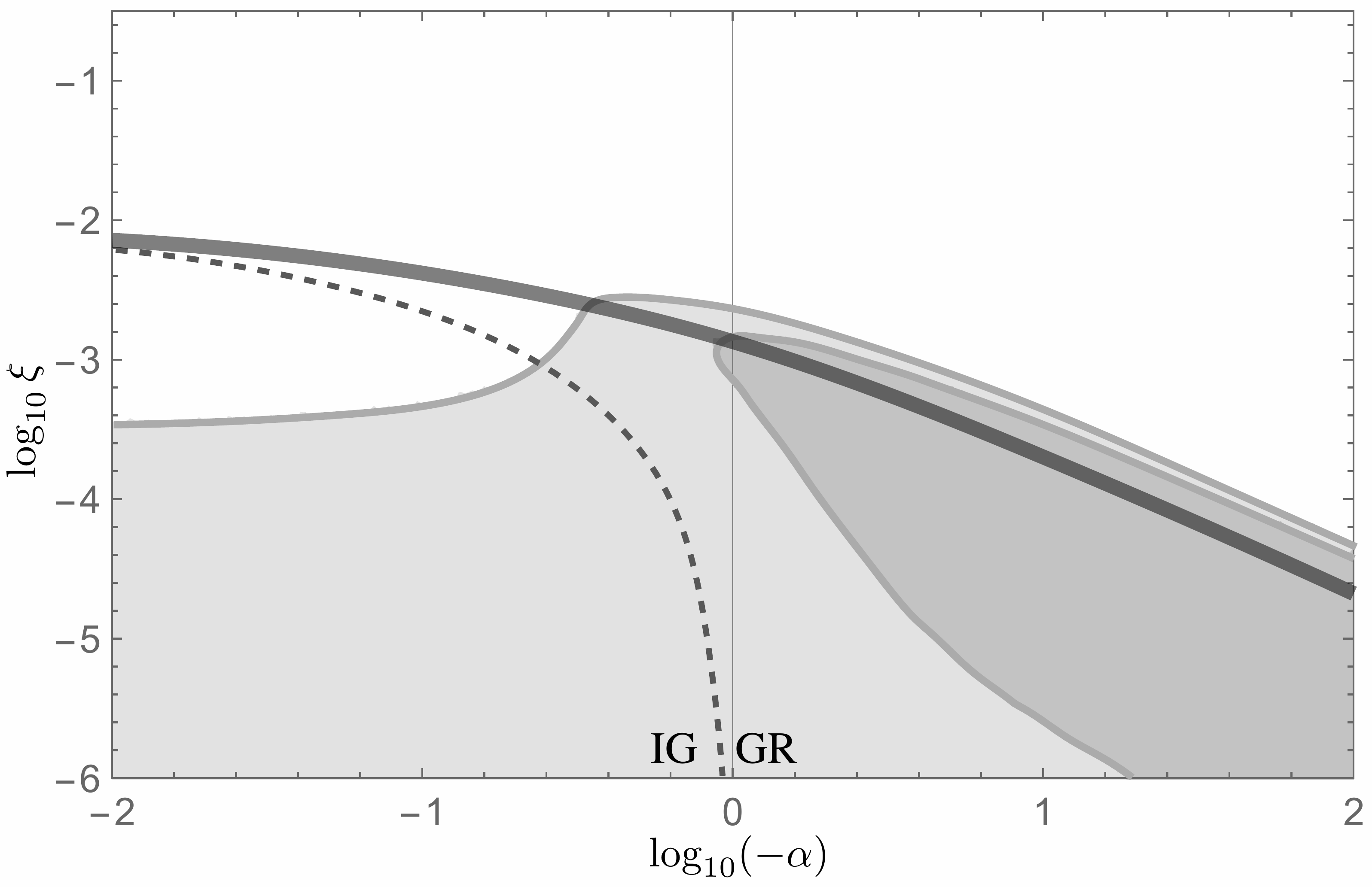, width=11. cm}
\caption{In this Figure, 3 different regions are shown. The darker grey region represents the set of $(\alpha,\xi)$ values leading to the theoretical predictions which lie inside the darker grey region in the $(n_s,r)$ plane of Figure (\ref{fig2}) (namely the $68\%$ confidence level region). Similarly the lighter grey region corresponds to the lighter grey part of the $(n_s,r)$ plane (namely the $95\%$ confidence level region) and finally the white area is that which leads to predictions outside the $95\%$ confidence level region. The dashed lines divide the parameters space in 2 regions corresponding to different inflationary regimes. The black band represents the subset of parameters corresponding to an evolution close to the fixed point (\ref{fp1}) for $k\sim \phi_*$.}
\label{fig3b}
\end{figure}

In this case the following $\M/k$ independent relation holds between $v_*$ and $u_*$:
\be{relfvV2}
v_*=\frac{\mu}{\alpha\xi}\frac{\pa{\rho_*+\alpha}^2}{\pa{\rho_*+1}^2}u_*^2
\ee
and on setting $k\sim\phi_*$ one finds that the linearity condition can be satisfied for $\xi$ small. In particular when $-\alpha\xi\ll 1$ the linearity condition for $u_*$ takes the form
\be{linV21}
\frac{\xi}{2}\pa{1-\alpha}\sim u_{FP}
\ee
while for $-\alpha\xi\gg 1$ ($N_*\sim 60$) and inflation occurring in the LF regime the same condition is $\xi\sim 10^{-2}$. Correspondingly the linearity condition for $v_*$ is satisfied and we find a subset of asymptotically safe inflationary models compatible with observations. \\
Conversely on setting $k\sim H_*$ and combining the relation $v_*\sim 6 u_*$ with (\ref{relfvV2}) one finds the result (\ref{fsV1}) which cannot satisfy the linearity condition since $u_{*}\gg 1$.\\
In Figure (\ref{fig2b}) we plotted a relevant part of the parameter space for the LF case and its comparison with observations. Let us note that, in contrast with the symmetric case, now the values of $-\alpha\ll 1$ are compatible with observations for $\xi\gg 1$. The parameter space is divided by a dashed line into 2 regions representing, on varying $\alpha$ and $\xi$, the possible regimes for which inflation takes place. The black band represents the subset of theories which satisfy the linearity conditions (\ref{linearcond}) on setting $k\sim \phi_*$.\\
In Figure (\ref{fig3b}) we plotted a relevant part of the parameter space for the SF case and its comparison with observations. The Figure shows that $\xi\ll 1$ is necessary in order to fit observations. In particular when $\xi\ll 1$, $-\alpha\gg 1$ there exists a subset of the parameter space with $-\alpha \xi\ll 1$ within the 68\% confidence level region of the $(n_s,r)$ plane. The parameter space is divided by a dashed line into 2 regions representing, on varying $\alpha$ and $\xi$, the possible regimes for which inflation takes place. The black band represents the subset of theories which satisfy the linearity conditions (\ref{linearcond}) on setting $k\sim \phi_*$. Let us finally note that, similarly to the symmetric case with $\xi\ll 1$, these models predict a tensor to scalar ratio $r$ of the order of the SR parameters.

\section{Conclusions}
In this article we studied how the theoretical constraints derived from AS could impact on inflationary model building. Our analysis focussed on a class inflationary models with a massive, self interacting scalar field non-minimally coupled to the Ricci scalar. On the phenomenological side this class of models can reproduce inflationary observations for a wide range of their parameters but, from the theoretical point of view, their dynamics must be tested against quantum corrections which, at planckian energy, can be relevant. In the absence of an universally accepted theory of quantum gravity which predicts the particle content of the very early universe and solves the problems related to the quantisation of the gravitational interaction, the AS scenario provides a valid alternative.\\
The set of theories lying on the critical surface associated with the fixed point (\ref{fp1}) are asymptotically safe and can be considered as candidates for inflation. Instead of randomly generating inflationary models - by spanning the space of initial conditions of the RG flow and then solving the non linear equations (\ref{flowvfull}), (\ref{flowufull}) - and checking, for each of them\footnote{in order to compare theoretical predictions with observations one must work out the analytical expressions for the observables for each different theory generated by the RG flow;}, their compatibility with inflationary observables we adopted a more convenient and model independent approach. We restricted our analysis to the class of theories (\ref{genform}) with $V$ and $U$ given by (\ref{renpot}), (\ref{rencoup}) \cite{ASgippo}. The form of these functions is generated by the RG flow in the linear approximation (close to the fixed point).  Only a subset of the parameter space leads to viable inflationary models. Then, on picking the models closest to the fixed point and restricting the parameter space accordingly, one finds an infinite set of the Asymptotically Safe inflationary models.\\
Our approach leads to the very interesting results shown in the figures. Let us note that the identification of the IR scale $k$ plays an important role in constraining the space of theories and we found non trivial results on setting $k\sim \phi_*$ where $\phi_*$ is the value of the inflaton field during inflation. In such a case AS puts severe constraints on the viable models. Even if observations do not exclude inflationary setups with a large non minimal coupling, $\xi$, once the requirement for AS is imposed, $\xi$ turns to be small and, for some regimes, close to $10^{-2}-10^{-3}$ with a slight dependence on shape of the potential parametrised by $\alpha$. Let us note that Higgs inflation, having $\xi\gg 1$, does not belong to this set of theories. In general, on varying $\alpha$ it is possibile to obtain viable inflation both in the IG and in the GR regimes. Finally let us note that AS gives distinctive predictions for the tensor to scalar ratio $r$: since $\xi\ll 1$ the tensorial spectral index is comparable with a first order SR parameter thus leading to $r\simeq 10^{-2}-10^{-1}$ but not too small as happens for Higgs inflation or, more generally, when $\xi\gg 1$. Different choices for the identification of $k$ lead to very different predictions. For example on setting $k\simeq H_*$ it is not possibile to satisfy the condition (\ref{linearcond}). However obeying to the linearity condition is a sufficient but not necessary condition to conclude that some model is closer to the critical surface. Consequently the models which do not satisfy the condition (\ref{linearcond}), and then lie far from the fixed point (\ref{fp1}), are not ruled out a priori and their position w.r.t the critical surface must be checked by means of the full RG flow. 

\section*{Acknowledgments}
I would like to thank Gian Paolo Vacca for illuminating discussions and useful comments and suggestions.


\end{document}